\documentclass[useAMS,usenatbib]{mn2e}

\usepackage{amsmath}
\usepackage{amssymb}
\usepackage{graphicx}% Include figure files
\usepackage{dcolumn}% Align table columns on decimal point
\usepackage{bm}% bold math
\usepackage{ulem}

\title[Thermodynamic Functions of Magnetized Coulomb Crystals]
{Thermodynamic Functions of Magnetized Coulomb Crystals}
\author[D. A. Baiko and D. G. Yakovlev]
{D. A. Baiko\thanks{E-mail:baiko@astro.ioffe.ru} and D. G. Yakovlev \\
A. F. Ioffe Physical-Technical Institute,
Politekhnicheskaya 26, 194021 St.-Petersburg, Russian Federation}

\begin{document}

\date{Accepted; Received ; in original form}

\pagerange{\pageref{firstpage}--\pageref{lastpage}} \pubyear{2013}

\maketitle

\label{firstpage}

\begin{abstract}
Free energy, internal energy, and specific heat for each of the three
phonon spectrum branches of a magnetized Coulomb crystal with
body-centered cubic lattice are calculated by numerical integration
over the Brillouin zone in the range of magnetic fields $B$ and
temperatures $T$, such that $0 \le \omega_{\rm B}/\omega_{\rm p}\le
10^3$ and $10^{-4} \le T/T_{\rm p} \le 10^4$. In this case,
$\omega_{\rm B}$ is the ion cyclotron frequency, $\omega_{\rm p}$ and
$T_{\rm p}$ are the ion plasma frequency and plasma temperature,
respectively. The results of numerical calculations are approximated
by simple analytical formulas. For illustration, these formulas are
used to analyze the behavior of the heat capacity in the crust of a
neutron star with strong magnetic field. Thermodynamic functions of
magnetized neutron star crust are needed for modeling various
observational phenomena in magnetars and high magnetic field pulsars.
\end{abstract}

\begin{keywords}
dense matter -- stars: neutron.
\end{keywords}

%%%%%%%%%%%%%%%%%%%%%%%%%%%%%%%%%%%%%%%%%%%%%%%%%%%%%%%%%%%%%%%%%%%%

\section{Introduction}
\label{introduct}
Coulomb crystals consist of fully ionized ions with charge $Ze$, mass
$M$, and number density $n$, arranged in a crystal lattice and
immersed into the electron background of constant and uniform density
$Zn$, which compensates the electric charge of the ions. The Coulomb
crystal model is used for description of such diverse physical
systems as dusty plasma and trapped ion plasma (e.g.,
\citealt{itano98,dubin99}), matter in white dwarf cores and neutron
star crusts (e.g., \citealt{HPY07}).

In astrophysics it is also important to study Coulomb crystals in the
presence of a magnetic field. This topic has gained relevance due to
the association of the most prominent gamma-ray sources, the
soft-gamma repeaters and anomalous X-ray pulsars (SGRs and AXPs) --
called collectively magnetars -- with isolated neutron stars possesing
extremely strong magnetic fields $B \gtrsim 10^{14}$~G
\citep[e.g.,][]{WT06,sandro08}. For instance, the
surface magnetic field of the SGR 1806--20, inferred from
measurements of its spin-down rate,
is\footnote{SGR/AXP~Online~Catalog: \\
http://www.physics.mcgill.ca/$\sim$pulsar/magnetar/main.html} $B \sim
2 \times 10^{15}$~G. Another example is the magnetar 1E 2259+586,
classified as an AXP, with the measured
spin-down field
$B \sim 5.9 \times 10^{13}$~G (e.g., \citealt{pp11}). In addition to
magnetars, there is a class of ``ordinary''\ pulsars which possess
very strong magnetic fields (as derived from their spin-down data).
Observations indicate that magnetars are most likely powered by
very strong
magnetic fields. In contrast, high magnetic field pulsars are powered
by magnetic braking. For example, we mention the X-ray pulsar
J1846--0258 ($B\approx 5 \times 10^{13}$~G) and the radio pulsar
J1718--3718 ($B\approx 7.5 \times 10^{13}$~G) -- see, e.g.,
\citet{livingstone11}, \citet{zhuetal11}, and references therein.

The measured
spin-down
magnetic fields of magnetars and high-$B$ pulsars are thus of
comparable strength but observational manifestations of these objects
are drastically different. Magnetars seem to be hotter and
demonstrate violent bursting activity, while high-$B$ pulsars are
quieter and colder. Interestingly though, observations of the
high-$B$ pulsar J1846--0258 showed (e.g., \citealt{livingstone11})
that in May-July of 2006 it demonstrated distinctly magnetar-like
X-ray bursts followed by a pulsar glitch and a return to the high-$B$
pulsar regime. Therefore, the evolution of high-$B$ pulsars and
magnetars can be related.

The internal fields of these stars can be much larger than their
spin-down surface
fields and can have both poloidal and toroidal components. Some
theoretical models of magnetars predict the internal crustal magnetic
fields in the range from a few $10^{14}$~G to a few $10^{16}$~G
(e.g., \citealt{pmg09,pp11} and references therein). This suggests
that all these stars possess the crust of highly magnetized Coulomb
crystals. One needs to study thermodynamics of these crystals to
model observational manifestations of magnetars, high-$B$ pulsars,
and their possible mutual transformations.

Magnetized Coulomb crystals have already been studied in a number of
works. We mention pioneering works of \citet[][]{UGU80} and
\citet[][]{NF82,NF83}; see \citet[][]{B09} for a summary of these
early results. In \citet[][]{B00,B09} a more quantitative study of
properties of the Coulomb crystal in the magnetic field has been
undertaken. In particular, the phonon mode spectrum of the crystal
with body-centered cubic (bcc) lattice has been calculated for a wide
range of magnetic field strengths and orientations. The phonon
spectrum has been used for a detailed analysis of the phonon
contribution to the crystal thermodynamic functions, Debye-Waller
factor of ions, and the rms ion displacements from the lattice nodes
for a broad range of densities, temperatures, chemical compositions,
and magnetic fields. The thermodynamic functions calculated by
\citet[][]{B00,B09} have been recently parameterized by \citet{pc13}.

In this paper we perform more extended calculations of the bcc
Coulomb crystal thermodynamic functions in the magnetic field. In
addition, we present analytical expressions which fit these numerical
results. They are more detailed and accurate than those presented by
\citet{pc13} but contain more fit parameters. We use them to analyze
the main features of the heat capacity of ions in a magnetized outer
and inner neutron star crust.

\section{General Theory}
\label{general}
The effect of the magnetic field $B$ on the ion
motion can be characterized by the ratio
\begin{equation}
   b = \omega_{\rm B} / \omega_{\rm p},
\label{rhoB}
\end{equation}
where
\begin{equation}
  \omega_{\rm B}=\frac{ZeB}{Mc}, \quad
  \omega_{\rm p}=\sqrt{\frac{4 \pi Z^2 e^2 n}{M}}
\end{equation}
are the ion cyclotron frequency and plasma frequency,
respectively, while $c$ is the speed of light.

Phonon frequencies $\Omega$ of a magnetized Coulomb cystal
are solutions of the following secular equation:
\begin{equation}
       {\rm det} \left\{ D^{\alpha \beta}(\bm{k}) -
\Omega^2 \delta^{\alpha \beta} - i \Omega \omega_{\rm B}
     \varepsilon^{\alpha \gamma \beta} n^\gamma  \right\} =0~.
\label{secular-2}
\end{equation}
In this case $\bm{n}$ is the unit vector in the direction of the magnetic field,
${\bm k}$ is the phonon wavevector in the first Brillouin zone (BZ),
and $D^{\alpha \beta}(\bm{k})$ is the dynamic matrix
of the lattice in the absence of the magnetic field.
Greek indices denote Cartesian coordinates $x,y,z$, and summation over repeated indices is assumed.
This matrix determines
frequencies $\omega_{\bm{k}j}$ and polarization vectors $\bm{e}_{\bm{k}j}$
of crystal oscillations at $\bm{B}=0$:
$D^{\alpha \beta} (\bm{k}) e^\beta_{\bm{k}j} = \omega^2_{\bm{k}j} e^\alpha_{\bm{k}j}$,
where $j$ enumerates oscillation modes with given $\bm{k}$
($j=1,2,3$).

We denote solutions of Eq.\ (\ref{secular-2}), i.e.\ magnetized
crystal phonon frequencies at a given $\bm{k}$, as $\Omega_{\bm{k}s}$
($s=1,2,3$). A detailed analysis of these modes for magnetized bcc
lattice was performed in \citet[][]{UGU80,NF82,NF83,B00,B09}. In summary,
there are three branches of the phonon spectrum, with the minimum,
intermediate, and maximum frequencies, which we denote
$\Omega_{1,2,3}$. If $k \to 0$ outside of the plane ${\bm k} \perp
{\bm B}$, $\Omega_1 \propto k^2$, while $\Omega_{2,3}$ tend to
constant values depending on the direction of ${\bm k}$. If $k \to 0$
in the plane ${\bm k} \perp {\bm B}$, $\Omega_{1,2} \propto k$, while
$\Omega_3$ tends to a constant value. At any ${\bm k}$ phonon
frequencies satisfy the generalized Kohn sum rule $\sum_s
\Omega_{\bm{k}s}^2 = \omega_{\rm p}^2 + \omega_{\rm B}^2$
\citep[][]{NF83}.

%******************************************************************
%                                                       spectrum
\begin{figure*}
%\vspace{-0.5cm}
\begin{center}
\leavevmode
\includegraphics[width=170mm,bb=16 18 684 343,clip]{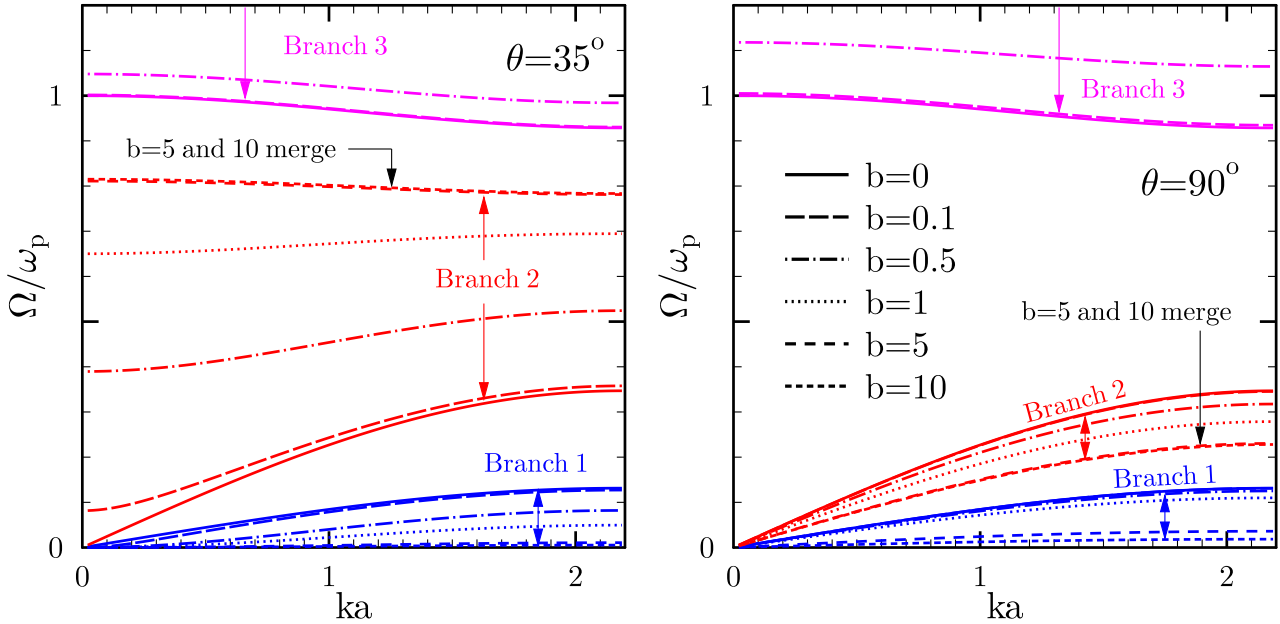}
\end{center}
\vspace{-0.4cm} \caption[ ]{(Color online) Phonon spectrum of the
magnetized bcc lattice for several values of $b = \omega_{\rm
B}/\omega_{\rm p}$. Magnetic field ${\bm B}$ and wavevector ${\bm k}$
orientations are defined in the text. Left and right panels
correspond to angles between ${\bm k}$ and ${\bm B}$ of $35^\circ$
and $90^\circ$, respectively. Branch 3 curves for $b \ge 1$ are not
shown. $a=(4 \pi n/3)^{-1/3}$ is the ion sphere radius.} \label{spec}
\end{figure*}
%
%******************************************************************

In this work we fix the direction of the magnetic field as ${\bm
n}=(1,1,1)/\sqrt{3}$. This direction coincides with the direction
towards one of the nearest neighbors in the bcc lattice. It also
corresponds to the minimum of the magnetized bcc crystal zero-point
energy \citep[][]{B09}. The phonon spectra in the directions of the
wavevector ${\bm k}/k=(1,1,0)/\sqrt{2}$ and ${\bm
k}/k=(1,-1,0)/\sqrt{2}$ are shown on the left and right panels of
Fig.\ \ref{spec}, respectively, for several values of $b$. Left panel
corresponds to an angle of about $35^\circ$ between ${\bm k}$ and
${\bm B}$. Right panel is for ${\bm k} \perp {\bm B}$.

Given the phonon frequencies, the phonon thermodynamic functions in
the magnetic field can be calculated using the same general formulas
\citep[e.g.,][]{LL80} as in the field-free case. The phonon free
energy (with phonon chemical potential $\mu=0$ and without zero-point
contribution) reads
\begin{eqnarray}
      F &=& T
   \sum_{\bm{k}s} \ln{\left[1-\exp{\left(-\frac{\hbar \Omega_{\bm{k}s}}{T}
        \right)}\right]}
\nonumber \\
        &=&  V T \sum_s \int_{\rm BZ} \frac{{\rm d} \bm{k}}{(2 \pi)^3}
        \ln{\left[1-\exp{\left(-\frac{\hbar \Omega_{\bm{k}s}}{T}\right)}
        \right]}~,
\label{Fph}
\end{eqnarray}
where $V$ is the volume, and the integral is over the first Brillouin zone.
The phonon internal thermal energy $E$ and heat capacity $C$
are then given by
\begin{eqnarray}
          E &=& F - T \left(\frac{\partial F}{\partial T}\right)_{\mu, V}
          = \sum_{\bm{k}s} \frac{\hbar \Omega_{\bm{k}s}}{e^{\hbar
            \Omega_{\bm{k}s}/T}-1}~,
\label{E/N} \\
          C &=& - T \left(\frac{\partial^2 F}{\partial T^2}\right)_{\mu, V}
             = \frac{1}{4T^2} \sum_{\bm{k}s}
          \frac{\hbar^2 \Omega_{\bm{k}s}^2}{{\rm sinh}^2(\hbar
            \Omega_{\bm{k}s}/2T)}~.
\label{ECph}
\end{eqnarray}

\section{Numerical Integration}
\label{numeric}
For accurate and fast numerical integration over the first BZ we have
developed a new integration scheme. The need to do this stemmed from
the cylindrical geometry imposed on the system by the magnetic field.
The typically used BZ integration scheme \citep[][]{AG81,B00,BPY01}
takes into account crystal geometry only.

The Brillouin zone of the bcc lattice is a rhombic dodecahedron
(convex polyhedron with 12 rhombic faces). It consists of 48
identical {\it primitive domains} of the form $k^x \ge k^y \ge k^z
\ge 0,~k^x+k^y \le 2 \pi/a_{\rm l}$, where $a_{\rm l}$ is the bcc
lattice constant, $na_{\rm l}^3=2$. 48 domains are obtained by 6
permutations of Cartesian coordinates in the above inequalities and
by reflections of 6 resulting domains with respect to planes $k^x=0$,
or $k^y=0$, or $k^z=0$, or any combination of the three for a total
of 8 distinct possibilities. Each rhombic face can be split by its
diagonals into 4 identical triangles (hereafter {\it primitive
triangles}) and each of the resulting 48 triangles is a face of the
respective primitive domain.

In this paper we calculate thermodynamic functions due to the three
branches of the phonon spectrum separately. In the absence of the
magnetic field it is sufficient to integrate in Eq.\ (\ref{Fph}) over
only one primitive domain as all domains yield identical
contributions. In the presence of the field it is no longer so.
Moreover, dominant contributions to the thermodynamic functions come
from different domains for different phonon branches. For instance,
for $k \to 0$ the intermediate branch $\Omega_2$ is acoustic
($\Omega_2 \propto k$) in the ${\bm k}\perp{\bm B}$ plane and is
optic ($\Omega_2 \to {\rm const}$) outside of this plane with
$\Omega_2(0)$ increasing with decrease of the angle $\theta$ between
${\bm k}$ and ${\bm B}$ (cf.\ Fig.\ \ref{spec}). At low temperatures
this results in $T^4$ dependence of the intermediate branch specific
heat with dominant contributions coming from the domains containing
the ${\bm k} \perp {\bm B}$ plane. The lowest branch $\Omega_1$ is
also acoustic for ${\bm k} \perp {\bm B}$ but is quadratic in $k$ as
$k \to 0$ for other angles between ${\bm k}$ and ${\bm B}$. This
results in $T^{3/2}$ dependence of the specific heat at low
temperatures with maximum contributions coming from domains
containing ${\bm k} \parallel {\bm B}$.

The integration over ${\bm k}$ proceeds as follows. For every phonon
branch we specify a $\theta$-grid ($\theta$ is the angle between
${\bm k}$ and ${\bm B}$), which is denser at those $\theta$, where
the maximum contribution to the thermodynamic functions of the given
branch is expected (e.g., around $\theta=\pi/2$ for branch
$\Omega_2$). For every primitive triangle we specify vertices $A$,
$B$, and $C$ corresponding to minimum, intermediate, and maximum
angles $\theta_{\rm min}$, $\theta_{\rm int}$, and $\theta_{\rm
max}$, respectively. These $\theta$ are added to the $\theta$-grid.
For every grid-point $\theta_i$ from the segments $[\theta_{\rm
min},\theta_{\rm int}]$, $[\theta_{\rm min},\theta_{\rm max}]$, and
$[\theta_{\rm int},\theta_{\rm max}]$ we find points on the edges of
the primitive triangle $[A,B]$, $[A,C]$, and $[B,C]$, respectively,
which are characterized by this $\theta_i$. For $\theta_{\rm min} <
\theta_i < \theta_{\rm max}$ there will be two such points, $P_i$ and
$Q_i$. For two neighboring grid-points $\theta_i$ and $\theta_{i+1}$,
such that $\theta_{\rm min} < \theta_i < \theta_{i+1} < \theta_{\rm
max}$, we can define two tetrahedra $O P_i Q_i P_{i+1}$ and $O
P_{i+1} Q_{i+1} Q_i$, where $O$ is the origin of the ${\bm k}$-space.
If one of the grid-points corresponds to a vertex (say, $\theta_{i+1}
= \theta_{\rm max}$ corresponds to $C$), we can define only one
tetrahedron (in this case $O P_i Q_i C$).

We then integrate over these tetrahedra  using a generalization of the method
of \citet[][]{AG81}. In particular, to integrate over a tetrahedron
$OPQR$ in ${\bm k}$-space we switch to variables $\xi$, $\eta$, and $\zeta$:
${\bm k} = \xi {\bm k}(P) + \xi \eta [{\bm k}(Q)-{\bm k}(P)]  +
\xi \eta \zeta [{\bm k}(R)-{\bm k}(Q)]$,
where, for instance, ${\bm k}(P)$ is the vector from the origin to the point $P$.
The Jacobian for this variable change is $\xi^2 \eta \det{\{K_i^\alpha\}}$, where $3 \times 3$
matrix $K_i^\alpha$, $i=1,2,3$, $\alpha=x,y,z$, is defined as $K^\alpha_1=k^\alpha(P)$,
$K^\alpha_2=k^\alpha(Q) - k^\alpha(P)$, $K^\alpha_3=k^\alpha(R)- k^\alpha(Q)$.
We integrate over $\xi$, $\eta$, and $\zeta$ from 0 to 1 using 8-point Gauss method.
At low $T$, we additionally split the interval $[0,1]$ for $\xi$ into a number
of subintervals $[0,\xi_0], [\xi_0,\xi_0^{(p-1)/p}], \ldots, [\xi_0^{1/p},1]$
with $\xi_0 = 10^{-6}$ and $p \sim 10$.

%%%%%%%%%%%%%%%%%%%%%%%%%%%%%%%%%%%%%%%%%%%%%%%%%%%%%%%%%%%%%%%%%%%%%%%%%%
\section{Results}
\label{s:fit}
%%%%%%%%%%%%%%%%%%%%%%%%%%%%%%%%%%%%%%%%%%%%%%%%%%%%%%%%%%%%%%%%%%%%%%%%%%%

\subsection{3-branch thermodynamics}
\label{T4}

%******************************************************************
%                                                       heatcap
\begin{figure*}
%\vspace{-0.5cm}
\begin{center}
\leavevmode
\includegraphics[width=170mm]{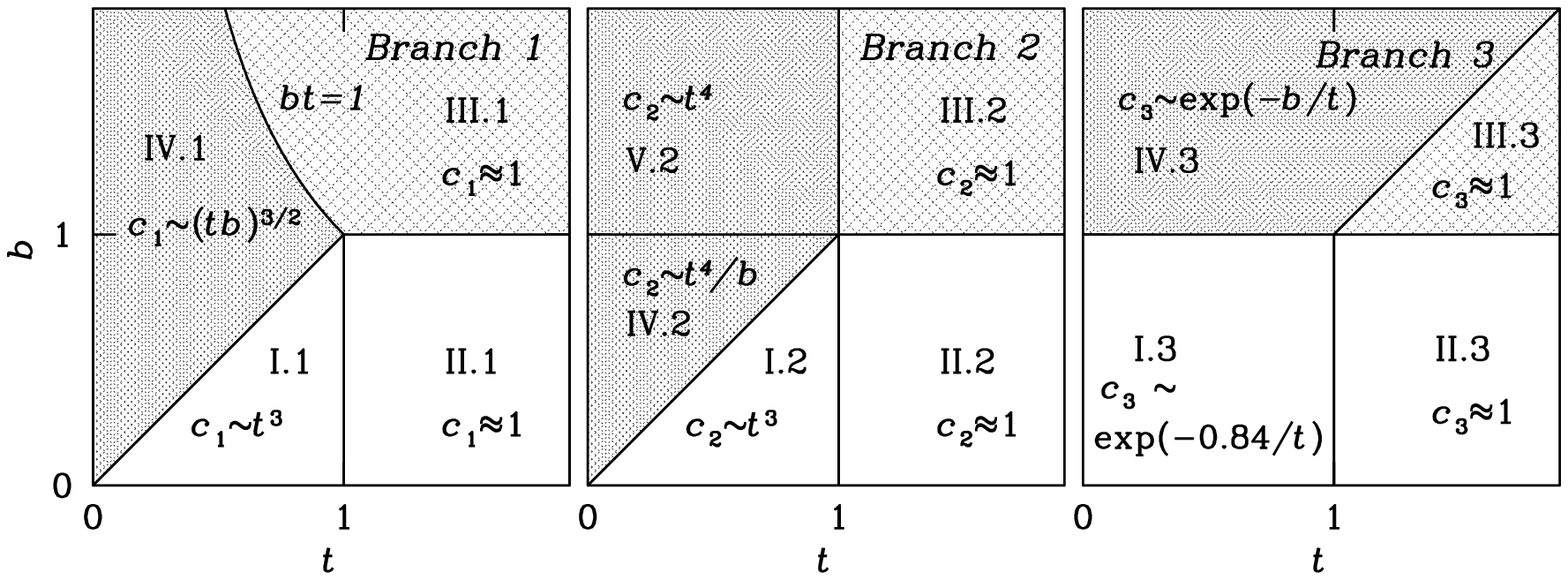}
%\end{center}
%\vspace{-0.4cm}
\caption{Schematic representation of regions of dimensionless
temperatures $t$ and magnetic fields $b$, where thermodynamic
quantities have different behavior, for branches 1, 2 and 3 (left,
middle, and right panels, respectively). Formulas give asymptotes of
partial heat capacities $c_i(b,t)$ in these regions. In densely
shaded regions (IV and V) magnetic field affects thermodynamics
significantly, while in open regions (I and II) it is unimportant. In
lightly shaded regions III the field affects only $f_i(b,t)$. See
text for details.} \label{f:modes}
\end{center}
\end{figure*}
%
%******************************************************************

The Helmholtz free energy $F$, internal thermal energy $E$,
and heat capacity $C$ of a magnetized Coulomb crystal can be
conveniently presented in the form
\begin{equation}
       F=-NTf(b,t),\quad E=NTe(b,t),\quad C=Nc(b,t),
\label{e:thermodyn}
\end{equation}
where $N$ is the number of ions; $f(b,t)$, $e(b,t)$, and $c(b,t)$ are
dimensionless functions of the dimensionless temperature and magnetic
field, $t=T/T_\mathrm{p}$ and $b=\omega_{\rm B}/\omega_{\rm p}$.

\begin{table}
\caption[]{Asymptotic regimes of thermodynamic functions}
\label{tab:modes}
\begin{center}
\begin{tabular}{c c c c  }
\hline \hline
Branch & Regime & $(b,t)$-range & Type \\
\hline
1 & I.1  & $b \ll t \ll 1$ & PL \\
1 & II.1 & $t \gg 1$ at $b\lesssim 1$ & SAT \\
1 & III.1 & $t \gg 1/b$ at $b\gtrsim 1$ & SAT \\
1 & IV.1 & $t \ll b$ at $b \lesssim 1$; $t \ll 1/b$ at $b \gtrsim 1$
& PL \\
\hline 2 & I.2  & $b \ll t \ll 1$ & PL \\
2 & II.2 & $t \gg 1 $ at $b \lesssim 1$& SAT \\
2 & III.2 & $t \gg 1 $ at $b \gtrsim 1$ & SAT \\
2 & IV.2 & $t \ll b$ at $b\lesssim 1$ & PL \\
2 & V.2 & $t \ll 1$  at $b\gtrsim 1$ & PL \\
\hline 3 & I.3 & $t \ll 1$ at $b \lesssim 1$ & EXP \\
3 & II.3 & $t \gg 1$ at $b\lesssim 1$ & SAT \\
3 & III.3 & $t \gg b$ at $b\gtrsim 1$ & SAT \\
3 & IV.3 & $t \ll b$ at $b\gtrsim 1$ & EXP \\
\hline \hline
\end{tabular}
\end{center}
\end{table}

The thermodynamic functions can be split into partial contributions
corresponding to the three phonon spectrum branches ($i=1,2,3$):
\begin{eqnarray}
  && f(b,t)=\sum_i f_i(b,t),\quad e(b,t)=\sum_i e_i(b,t),\nonumber \\
  && c(b,t)=\sum_i c_i(b,t).
\label{e:thermodyn_i}
\end{eqnarray}
These partial contributions describe three thermodynamics, each for a
particular branch~$i$. The combined thermodynamic quantities
must satisfy the Bohr-van Leeuwen theorem, which states
that thermodynamic functions of a classic system do not
depend on magnetic field. Our system is classic when occupation
numbers of $all$ phonon modes are large.

\begin{table}
\caption[]{Asymptotic behavior of heat capacity fits in PL regimes}
\label{tab:pl}
\begin{center}
\begin{tabular}{c c c c  }
\hline \hline
Branch $i$ & Regime & PL index $\gamma$& Asymptote of $c_i(b,t)$ \\
\hline
1 & I.1  & 3 & $2026\, t^3$ \\
1 & IV.1 & 1.5 &
$31.1 (bt)^{3/2}$ \\
\hline 2 & I.2  & 3 & $486.2\, t^3$ \\
2 & IV.2 & 4 & $1572 t^4 b^{-1}$ \\
2 & V.2 & 4 & $5019 t^4$ \\
\hline \hline
\end{tabular}
\end{center}
\end{table}

For any given $i$ the functions $f_i(b,t)$, $e_i(b,t)$, and
$c_i(b,t)$ are related through [cf.\ Eqs.\ (\ref{Fph})--(\ref{ECph})]
\begin{equation}
    e_i(b,t)=t \, \frac{\partial f_i(b,t)}{\partial t},\quad
    c_i(b,t)=\frac{\partial\, t e_i(b,t)}{\partial t}.
\label{e:diff}
\end{equation}
Accordingly, it is sufficient to calculate $f_i(b,t)$ to determine
the other functions.

\begin{table}
\caption[]{Asymptotic behavior of free energy fits in SAT regimes}
\label{tab:sat}
\begin{center}
\begin{tabular}{c c c  }
\hline \hline
Branch $i$ & Regime &  Asymptote of $f_i(b,t)$ \\
\hline
1 & II.1  &  $\ln t + 1.392$ \\
1 & III.1  &  $\ln (tb) + 1.795 $ \\
\hline 2 & II.2  &  $\ln t + 0.9338$ \\
2 & III.2  &  $\ln t +0.70 $ \\
\hline 3 & II.3  &  $\ln t +0.168 $ \\
3 & III.3  &  $\ln (t/b) $ \\
\hline \hline
\end{tabular}
\end{center}
\end{table}

It is convenient to distinguish different regimes of thermodynamic
functions. For a magnetized crystal there are a total of 13 regimes for
the three phonon branches. They are realized in different regions of $b$ and $t$
sketched in Fig.\ \ref{f:modes} and listed in Table
\ref{tab:modes}. These thermodynamic regimes can be divided into
low-temperature (quantum) and high-temperature (classic) ones.

In the low-temperature regimes, only a small fraction of phonon modes has
non-zero occupation numbers, so that the thermodynamic
functions are small [$f_i(b,t)\ll 1$, $e_i(b,t)\ll 1$, $c_i(b,t)\ll
1$]. The low-temperature regimes are further
subdivided into those where the thermodynamic functions decrease with
decreasing temperature according to a power law
or exponentially (PL or EXP regimes in Table \ref{tab:modes}).
In general, we have 5 quantum PL regimes (I, IV, V for branches 1 and 2)
and two quantum EXP regimes (I and IV for branch 3 only).

According to Eq.\ (\ref{e:diff}),
in a PL regime with index $\gamma$ we have
\begin{equation}
 c_i(b,t)= (\gamma+1) e_i(b,t)= (\gamma+1)\gamma f_i(b,t)=c_{i0}t^\gamma, \label{e:pl}
\end{equation}
where $c_{i0}$ is independent of $t$ but may depend on $b$. The
specific PL asymptotes of the heat capacity are listed in Table
\ref{tab:pl}. These expressions follow from our fit expressions
(Sects.\ \ref{fit0} and \ref{fitB}) and are consistent with our
numerical results. We see that in the 5 PL regimes we have
$\gamma=$3, 3/2, and 4. The PL regime with $\gamma=3$ (regime I.1 or
I.2) is the textbook Debye case of $T^3$ dependence of the heat
capacity due to acoustic phonons. The PL dependence of magnetized
crystal heat capacity with $\gamma=3/2$ (regime IV.1) was predicted
by \citet[][]{UGU80} based on the $k^2$ dependence of the branch 1
frequency at small $k$. Similar dependence is obtained for magnon
heat capacity \citep[e.g.,][]{K95}.

The $T^4$ regime reported here for branch 2 heat capacity (regimes
IV.2 and V.2) stems from the combined dependence of the frequency on
$k$ and on the angle $\chi = \pi/2-\theta$ between the wavevector and
the plane orthogonal to the magnetic field. The behavior of the
frequency $\Omega_2$ at small $k$ and $\chi$ can be understood if one
reverses signs in front of the square roots in Eq.\ (14) or (15) of
\citet[][]{B09}. In both ways one obtains $\Omega_2 \approx
\omega_{\rm p} \sqrt{|E_B|/(1+b^2)}$, where $E_B$ is given by Eq.\
(12) of the same paper. To lowest order in $k$ and $\chi$, one can
then write
\begin{equation}
   \hbar \Omega_2 \approx \sqrt{\alpha^2 k^2 + \beta^2 \chi^2}~,
\label{O2a}
\end{equation}
where coefficients $\alpha^2$ and $\beta^2$ depend on $b$,  on the
azimuthal angle $\phi$ of ${\bm k}$ with respect to ${\bm B}$, and on
the direction of  ${\bm B}$; these coefficients are always positive.
Internal energy $E_2$ of the second branch phonons can be written as
[cf.\ Eqs.\ (\ref{Fph}) and (\ref{E/N})]
\begin{eqnarray}
           && \frac{ (2 \pi)^3 E_2}{V} = \int
           \frac{\hbar \Omega_{{\bm k}2} \,\, {\rm d}{\bm k}}
           {\exp{\left(\hbar \Omega_{{\bm k}2}/T\right)} - 1}
\nonumber \\
           && =\int_0^{2 \pi} {\rm d}{\phi} \int^{\pi/2}_{-\pi/2}
           {\rm d} \chi \cos{\chi} \int
           \frac{\hbar \Omega_{{\bm k}2}  k^2 \, {\rm d}k}
            {\exp{\left(\hbar \Omega_{{\bm k}2}/T\right)} - 1}~.
\label{E2}
\end{eqnarray}
In the $T \to 0$ limit one can extend integrations over $k$ and
$\chi$ out to infinity, replace $\cos{\chi}$ by 1 and employ Eq.\
(\ref{O2a}). After that, at each $\phi$ one can replace integration
variables $(k,\chi) \to (r,s)$, where $\alpha k = r \cos{s}$, $\beta
\chi = r \sin{s}$, and $r$ varies from 0 to $\infty$, while $s$
varies from $-\pi/2$ to $\pi/2$. This yields
\begin{equation}
       E_2 = \Phi \int_{-\pi/2}^{\pi/2} {\rm d}s \int_0^\infty {\rm d}r\,
       \frac{r^4 \cos^2{s}}{\exp{(r/T)}-1} =
       12 \pi \zeta(5) \Phi T^5,
\label{ET5}
\end{equation}
where $\Phi$ is a constant determined by the $\phi$-integral.
Heat capacity is then proportional to $T^4$.

In the high-temperature regimes (denoted as SAT in Table \ref{tab:modes}),
all phonon modes of a given branch are fully
excited. Partial internal energy and heat capacity functions
$e_i$ and $c_i$ saturate at their maximum value 1,
while free energies reach logarithmic asymptotes:
\begin{equation}
   e_i(b,t) \to 1, \quad c_i(b,t) \to 1, \quad f_i(b,t) \to
   \ln t+ F_i~.
\label{e:saturation}
\end{equation}
There are 6 classical (SAT) regimes (two regimes, II and III, for
each mode). In Eq.\ (\ref{e:saturation}), $F_i$ are
temperature-independent but can depend on $b$ (being determined by
the logarithm of a phonon frequency, $\ln \Omega_i$, averaged over
the Brillouin zone). At the same time, $\sum_i F_i$ is
$b$-independent in accordance with the Bohr-van Leeuwen theorem.
Values of $F_i$ used in our fits are given in Table \ref{tab:sat}.
They are consistent with asymptotes that can be extracted from our
numerical data.

Densely shaded ($b,t$)-regions in Fig.\ \ref{f:modes} show those
regimes where all thermodynamic functions (for a given branch) are
strongly affected by the magnetic field. These are seen to be quantum
regimes IV and V. Lightly shaded regions (regions III for any $i$) indicate the
saturation regimes where only $f_i(b,t)$ is affected by the magnetic
field [through $F_i$ in Eq.~(\ref{e:saturation})]. Finally, in blank
regions (I and II for any $i$) magnetic field effects are weak.

\begin{table}
\caption[]{Fit accuracy of thermodynamic functions}
\label{tab:errors}
\begin{center}
\begin{tabular}{c c c c c }
\hline \hline
Function & $\delta_\mathrm{rms}$ & $\delta_\mathrm{max}$ & $b_m$ & $t_m$ \\
\hline
$f(0,t)$ & 0.0022  & 0.0070 & 0 & 0.00501  \\
$e(0,t)$ & 0.0026  & 0.0081 & 0 & 0.0158 \\
$c(0,t)$ & 0.0030  & 0.0105 & 0 & 0.0126 \\
\hline
$f(b,t)$ & 0.019 & 0.106  & 0.00158 & 0.000126 \\
$e(b,t)$ & 0.019 & 0.113  & 0.00251 & 0.000126 \\
$c(b,t)$ & 0.019 & 0.117  & 0.00251  & 0.000100 \\
\hline \hline
\end{tabular}
\end{center}
\end{table}

\subsection{Calculations and fits}

Using the technique described in Sect.\ \ref{numeric} we have
calculated  9
thermodynamic functions $f_i(b,t)$, $e_i(b,t)$, and $c_i(b,t)$
defined by Eqs.\ (\ref{e:thermodyn}) and (\ref{e:thermodyn_i}). The
calculations were done on a dense grid of temperatures (81 $t$-points
logarithmically equidistant in the interval from $10^{-4}$ to
$10^{4}$). We have considered the non-magnetized crystal ($b=0$) and
31 magnetic field values $b$ (logarithmically equidistant in the
range from $10^{-3}$ to $10^{3}$).

The thermodynamic functions vary over many orders of magnitude in a
non-trivial manner. To facilitate the use of these results, we have
fitted the calculated functions by analytic expressions. The fit
formulas are presented and discussed below. Here we outline their
general features.

For branches $i=1$ and 2 we fit $f_i(b,t)$ by the functions
\begin{equation}
     f_i(b,t)=\ln \left( 1+ \frac{P_i(b,t)}{Q_i(b,t)}  \right),
\label{e:FIT12}
\end{equation}
where $P_i(b,t)$ and $Q_i(b,t)$ are certain sums of power-law
functions of $t$.

The cyclotron branch $3$ is different. Its thermodynamics is well
described by the Einstein model with some oscillator frequency
$\omega=\kappa(b)\, \omega_\mathrm{p}$:
\begin{equation}
    f_3(b,t)=\ln \left( \frac{1}{1-\exp(-\kappa(b)/t)}
    \right),
\label{e:FIT3}
\end{equation}
where $\kappa(b)$ is temperature independent.

We will present the analytic fit expressions for $f_i(b,t)$ only.
Other thermodynamic quantities can be calculated from $f_i(b,t)$
using Eq.\ (\ref{e:diff}). We have verified that the expressions for
$e_i(b,t)$ and $c_i(b,t)$, obtained from our analytic fits to
$f_i(b,t)$, describe well the calculated values of $e_i(b,t)$ and
$c_i(b,t)$. The asymptotes presented in Tables \ref{tab:pl} and
\ref{tab:sat} correspond to the fit expressions.

Table \ref{tab:errors} summarizes the accuracy of our fits. Three
upper lines give root-mean-square (rms) relative deviations
$\delta_\mathrm{rms}$ and maximum deviations $\delta_\mathrm{max}$ of
the fitted and calculated total thermodynamic functions (summed over
all branches) for a non-magnetized crystal ($b=0$). Deviations are
determined over all 81 temperature grid points. The fits are seen to
be rather accurate, with $\delta_\mathrm{rms}\lesssim 0.3\%$ and
$\delta_\mathrm{max}\lesssim 1\%$. Three last lines list
$\delta_\mathrm{rms}$ and $\delta_\mathrm{max}$ of the total
thermodynamic functions for a magnetized crystal (over all $81\times
32=2592$ grid points of $t$ and $b$). These fits are less accurate
($\delta_\mathrm{rms}\lesssim 2\%$ and $\delta_\mathrm{max}\lesssim
11\%$). The last two columns of Table \ref{tab:errors} present the
values of $b_m$ and $t_m$, where the maximum errors occur.

Note that recently \citet{pc13} have produced analytic fits to
thermodynamic functions calculated by \citet{B00,B09}. Our fit
expressions are different. We fit more extended set of more precisely
calculated numerical data. In addition, we approximate separately the
contributions of different phonon branches while \citet{pc13} fitted
the thermodynamic functions summed over phonon branches. Their fits
contain fewer fit parameters but are somewhat less accurate. For
instance, comparing those fits with our newly calculated $c(t,b)$ we
obtain $\delta_\mathrm{rms}\approx 12$\%, and
$\delta_\mathrm{max}\approx 56$\% (at $b_m=63.096$ and $t_m=0.0001$);
cf.\ Table \ref{tab:errors}.

\subsection{Field-free case}
\label{fit0}

Thermodynamics of non-magnetized Coulomb crystals is well studied.
The Debye temperature of the crystal can be shown to be
$\Theta=0.4532\, T_\mathrm{p}$ \citep[][]{C61}. For all three
branches one has two asymptotic regimes, the quantum regime $t\ll 1$,
and the classic one $t \gg 1$ (Fig.\ \ref{f:modes}: I and II).

\begin{table}
\caption[]{Fit parameters  in Eqs.~(\ref{e:f120}),
(\ref{e:f1b})--(\ref{e:betafit})} \label{tab:a12}
\begin{center}
\begin{tabular}{c c c c c c c }
\hline \hline
$\ell$ & $a_{1\ell}$ &  $a_{2\ell}$ & $q_\ell$ & $r_\ell$ & $\mu_\ell$ & $\nu_\ell$ \\
\hline
1 & 168.83  & 40.517 & 2.569 & 4.503 & 2.737 & 1.650 \\
2 & 0.3317  & 0.0351 & --1.986 & 4.927 & --45100  & 0.2655 \\
3 & 122200  & 4718.4 & 0.5184 & 2.3156 & 25.09 & 1 \\
4 & 3.153  & 1.621  & 35.73 &  6.579 & 0  & 1 \\
5 & 316.7  & 65.82 & 1.162 & 1.570 & 37.15  & 0.8881 \\
6 & 3339  & 335.6  & 0.4856 &  2.252 & 31.26 & 1 \\
7 & 30370  & 1854.6 & 0.5640 & 2.335 & 32.01 & 1 \\
 \hline \hline
\end{tabular}
\end{center}
\end{table}

For branches $i=1$ and 2 we suggest the fits (\ref{e:FIT12}) with
\begin{eqnarray}
    P_i(0,t) &=& a_{i1} t^3+a_{i2}t^4+a_{i3}t^5,
\nonumber \\
    Q_i(0,t) &=& 1+a_{i4}t+a_{i5}t^2+a_{i6}t^3+a_{i7}t^4.
\label{e:f120}
\end{eqnarray}
The fit parameters $a_{i\ell}$ are collected in Table \ref{tab:a12}.
The quantum regimes I.1 and I.2 are the Debye power-law with
$\gamma=3$ in (\ref{e:pl}) due to the excitation of acoustic phonons
with frequencies $\Omega_{1,2} \ll \omega_\mathrm{p}$.

For the optical branch 3 we have Eq.\ (\ref{e:FIT3}) with
\begin{equation}
    \kappa=\kappa(0)=0.8443.
\label{e:f30}
\end{equation}
Accordingly, its contribution is exponentially suppressed at $t \ll
1$ in regime I.3 [$f_3(0,t) \sim e_3(0,t) \sim c_3(0,t) \sim
\exp(-\kappa/t) \ll 1$], but the saturation regime II.3 is similar to
those of the acoustic modes 1 and 2. Strictly speaking, at very low
$t\lesssim 0.01$ the calculated thermodynamic quantities for branch 3
start to deviate from the pure Einstein model (\ref{e:FIT3}), but
they become so small that their contribution to thermodynamics is
negligible. Therefore, the use of Eq.~(\ref{e:f30}) for all $t$ is
well justified.

The fits (\ref{e:f120}) and (\ref{e:f30}) are reasonably accurate
(Table \ref{tab:errors}). Note that more accurate fits to the total
functions $f(0,t)$, $e(0,t)$, and $c(0,t)$ are given by \cite{BPY01}.

Our calculations and fits reproduce (Table \ref{tab:pl}) the well
known low-temperature asymptote \citep[][]{C61}:
\begin{equation}
   c(0,t) =  2513 \,t^3\quad{\rm at}~t\ll 1~,
\label{e:asy:c0}
\end{equation}
and the well known high-temperature asymptote \citep[][]{PH73}:
\begin{equation}
   f(0,t) = 3(\ln t + 0.8313)\quad{\rm at}~t\gg 1~.
\label{e:asy:f0}
\end{equation}

\subsection{Magnetized crystal}
\label{fitB}

For the free energy function $f_1(b,t)$ of the magnetized crystal we
suggest the fit (\ref{e:FIT12}) with
\begin{eqnarray}
   P_1=& 8.293(bt)^{3/2}+
    a_{11}\alpha_1^3 t^3+a_{12}\alpha_2^4 t^4+a_{13}\alpha_3^5t^5,
\nonumber \\
   Q_1 =& 1+a_{14}\alpha_4t+a_{15}\alpha_5^2t^2
    +a_{16}\alpha_6^3t^3+a_{17}\alpha_7^4t^4,
\label{e:f1b}
\end{eqnarray}
where
\begin{equation}
\alpha_\ell=\sqrt{1+q_\ell b+r_\ell b^2},
\label{e:coeff_f1b}
\end{equation}
and the fit coefficients $q_\ell$ and $r_\ell$ are given in Table
\ref{tab:a12}.

This fit incorporates all the asymptotic regimes shown in Fig.\
\ref{f:modes}. In particular, it reduces to (\ref{e:f120}) in the
limit of $b \to 0$; it reproduces thermodynamic function slope change
(from $\gamma=3$ to $3/2$) in quantum regime IV as well as the
reduction of the branch 1 saturation temperature and the modification
of the classic asymptote of $f_1$ at $b \gg 1$ in regime III.

For branch 2 we obtain the fit (\ref{e:FIT12}) with
\begin{eqnarray}
   P_2&=&
    a_{21}\beta_1 t^4+a_{22}\beta_2 t^5+a_{23}\beta_3 t^6,
\nonumber \\
   Q_2& =& \beta_*+\beta_{**}\sqrt{t}+ t\nonumber \\
   & +& a_{24}\beta_4t^2+a_{25}\beta_5 t^3
    +a_{26}\beta_6 t^4+a_{27}\beta_7 t^5,
\label{e:f2}
\end{eqnarray}
where all magnetic field effects are included into the fit
coefficients
\begin{eqnarray}
  \beta_\ell&=&1+\frac{\mu_\ell \, b^2}{\nu_\ell+b^2}, \label{e:betafit} \\
%cofb6:=b/sqrt(1+b*b)*par[3]+add6*ff5
  \beta_* &=&  0.5154 \left(b^2 \over 1+b^2 \right)^{1/2}+0.08791\,
 \frac{b^2}{1+b^2},\nonumber\\
% cofb7:=sqrt(b/sqrt(1+b*b))*par[4]+add7*ff8
  \beta_{**}&=& -0.3527 \left(b^2 \over 1+b^2 \right)^{1/4}
   - 0.0706 \, \frac{b^2}{0.2655+b^2}.
\label{e:a16}
\end{eqnarray}
The fit coefficients $\mu_\ell$ and $\nu_\ell$ in Eq.\
(\ref{e:betafit}) are given in Table \ref{tab:a12}. In the limit
$b\to 0$ the coefficients $\beta_*$ and $\beta_{**}$ vanish and the
ratio $Q_2/P_2$ given by Eq.~(\ref{e:f2}) reduces to that given by
the field-free fit (\ref{e:f120}). By construction, the fit
(\ref{e:f2}) also reproduces the asymptotes of Fig.\ \ref{f:modes}
including the thermodynamic function slope change from $\gamma=3$ to
4 in quantum regimes IV and V and modification of the classic
asymptote of $f_2$ at large $b$ in regime III. At $b \gtrsim 1$
thermodynamics of branch 2 becomes independent of $b$ because the
spectrum of branch 2 phonons becomes field-independent (cf.\ Fig.\
\ref{spec}). This remarkable property is implanted in our fit
(\ref{e:f2}): all coefficients $\beta_\ell$, $\beta_*$ and
$\beta_{**}$ become $b$-independent at $b \gtrsim 1$.

%******************************************************************
%                                                       heatcap
\begin{figure}
%\vspace{-0.5cm}
\begin{center}
\leavevmode
\includegraphics[bb=20 40 515 625, width=84mm]{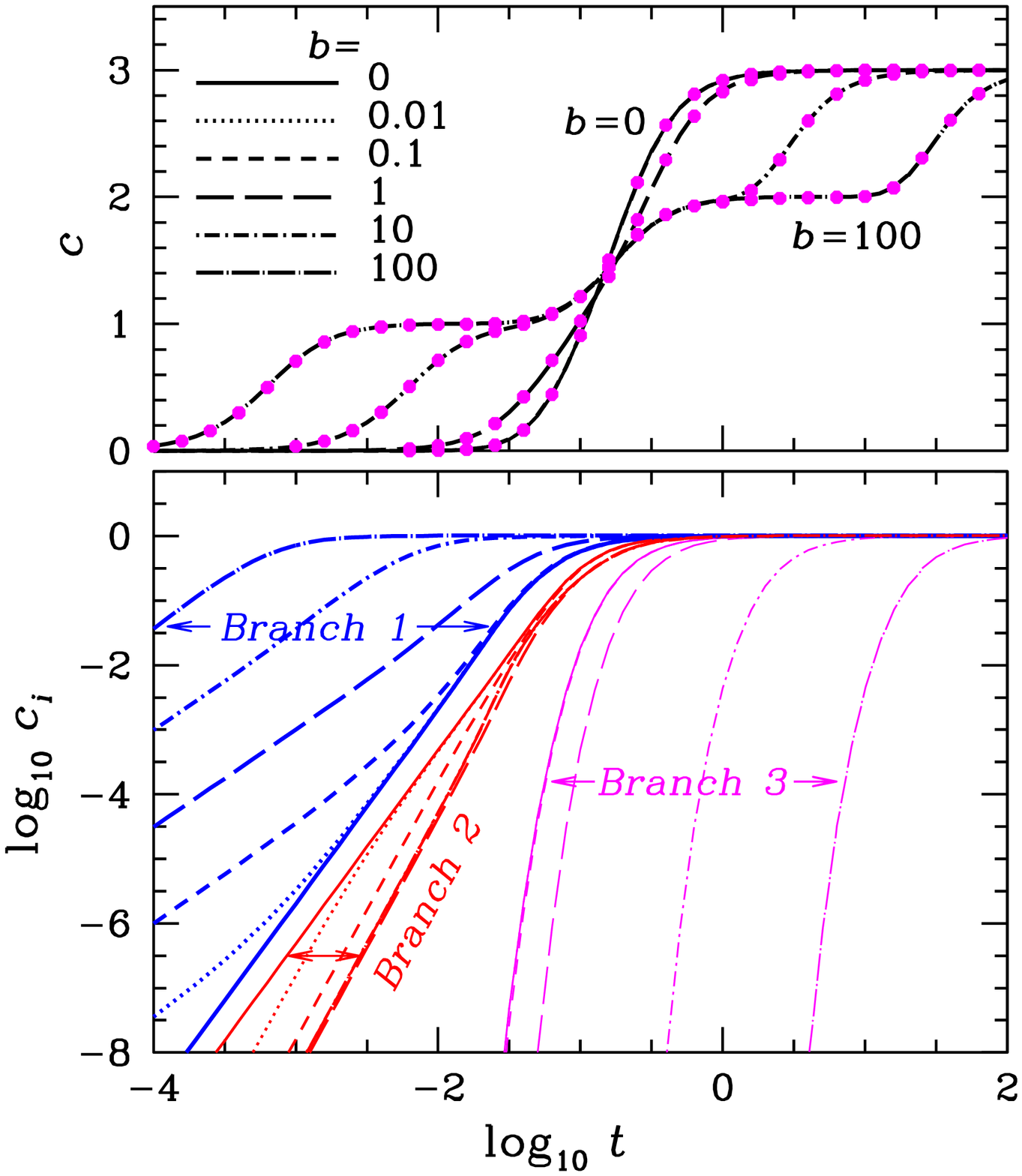}
\end{center}
\vspace{-0.4cm} \caption[ ]{(Color online) {\it Top:} Total heat
capacity (in linear scale) versus ${\rm log}_{10}t$ for $b=$0, 0.01,
0.1, 1, 10, and 100 (lines of different types are plotted using
original calculations; dots are given by the fit expressions at some
selected $t$ points). {\it Bottom:} Logarithm of partial heat
capacities $c_i(b,t)$ versus ${\rm log}_{10}t$ at the same $b$. See
text for details.} \label{f:capacity}
\end{figure}
%
%******************************************************************

%******************************************************************
%                                                       energy
\begin{figure}
%\vspace{-0.5cm}
\begin{center}
\leavevmode
\includegraphics[width=84mm]{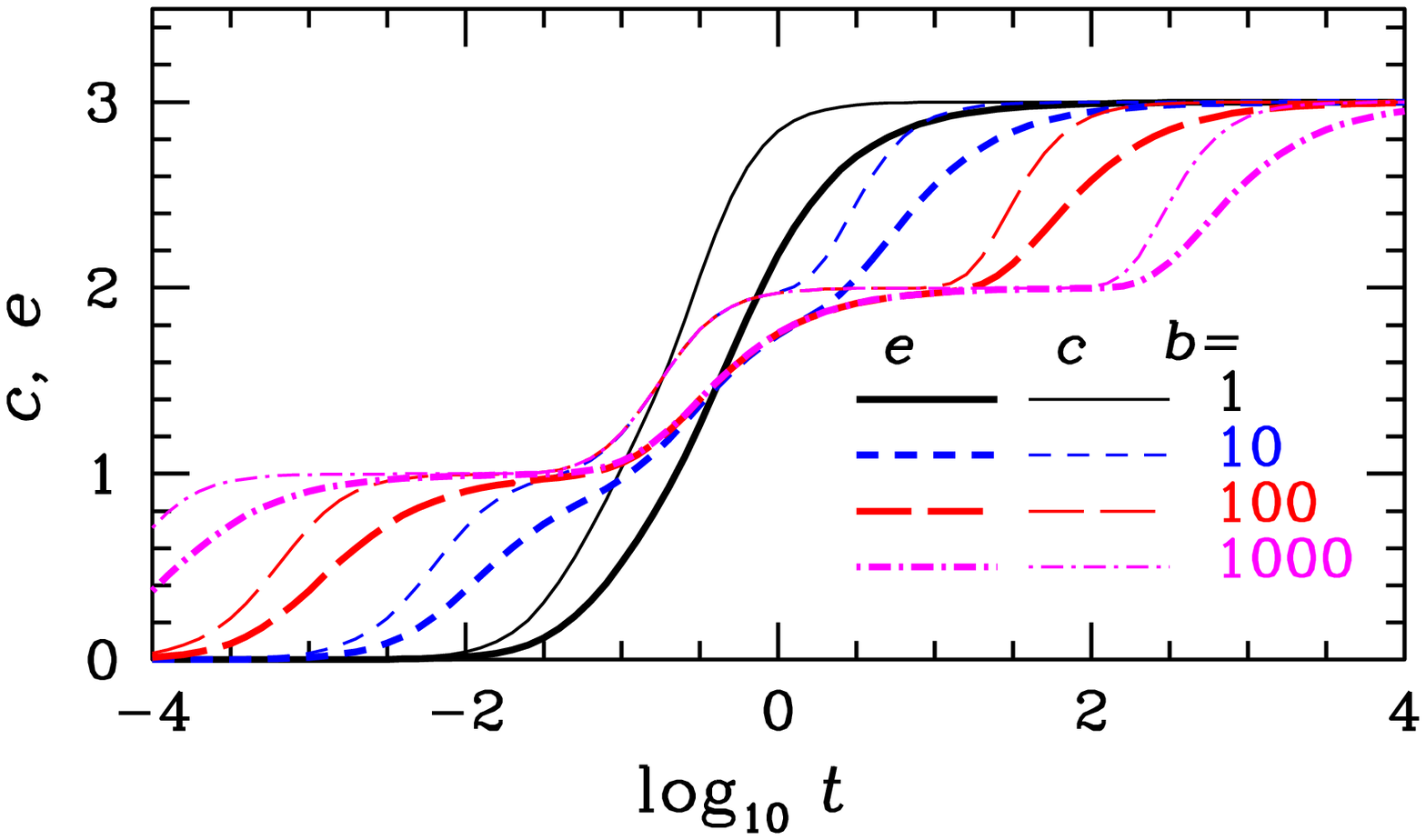}
\end{center}
\vspace{-0.4cm} \caption[ ]{(Color online) Total heat capacity
$c(b,t)$ (thin lines) and internal energy $e(b,t)$ (thick lines)
versus ${\rm log}_{10}t$ at $b=$1, 10, 100, and 1000 (lines of
different types). See text for details.} \label{f:energy}
\end{figure}
%
%******************************************************************

%******************************************************************
%                                                       entropy
\begin{figure}
%\vspace{-0.5cm}
\begin{center}
\leavevmode
\includegraphics[width=84mm]{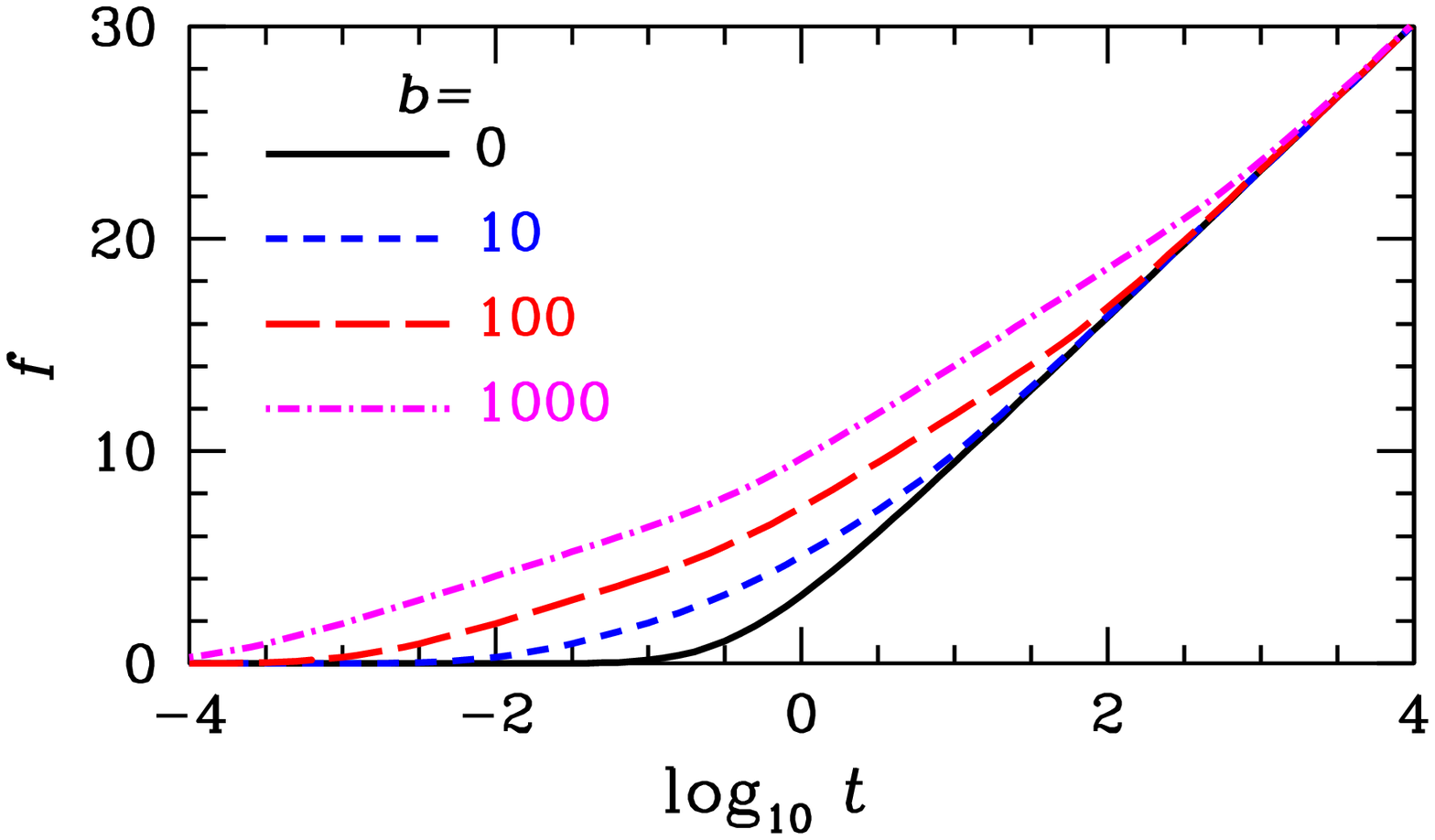}
\end{center}
\vspace{-0.4cm} \caption[ ]{(Color online) Total free energy $f(b,t)$
versus ${\rm log}_{10}t$ at $b=$0, 10, 100, and 1000 (lines of
different types). See text for details.} \label{f:entropy}
\end{figure}
%
%******************************************************************

Finally, the branch 3 thermodynamics is well described by the
Einstein model of one oscillator with the frequency
$\Omega_3=\sqrt{(0.8443\, \omega_\mathrm{p})^2+\omega_\mathrm{B}^2}$.
This implies Eq.~(\ref{e:FIT3}) with
\begin{equation}
     \kappa(b)=\sqrt{0.8443^2+b^2}~.
\label{e:f3}
\end{equation}
In the limit of $b \to 0$ this equation reduces to (\ref{e:f30}). In
the limit of $b \gg 1$ we have $\Omega_3 \to \omega_\mathrm{B}$,
which yields the thermodynamics due to cyclotron rotation of ions.
Again, expression (\ref{e:f3}) reproduces all asymptotic regimes
shown in Fig.\ \ref{f:modes} and Table \ref{tab:sat}. Because of the
Einstein character of the branch 3 spectrum, thermodynamic functions
in quantum regimes I.3 and IV.3 are exponentially suppressed.  There
are small deviations from the Einstein model in these regimes, but
these deviations can be neglected in the total thermodynamic
functions.

\subsection{Total thermodynamic functions}

Thermodynamics of magnetized Coulomb crystals was analyzed by
\cite{B00,B09}. We outline the main properties emphasizing the
contribution of different phonon branches into total thermodynamic
functions (\ref{e:thermodyn}). All the curves presented in this
section are plotted using calculated data. The fits would give very
close curves. To illustrate this point in the upper panel of Fig.\
\ref{f:capacity} we show the fit results for some selected values of
$b$ and~$t$.

Figure \ref{f:capacity} plots the heat capacity. The lower panel
shows the partial heat capacities $c_1(b,t)$, $c_2(b,t)$, and
$c_3(b,t)$ (on logarithmic scale) versus ${\rm log}_{10}t$ for $b=0$,
0.01, 0.1, 1, 10, and 100. When the magnetic field increases,
$c_1(b,t)$ grows up. In this way the magnetic field lowers the
saturation temperature (at which $c_1(b,t) \approx 1$) and the
power-law index (from 3 to 3/2) in quantum regime (Table
\ref{tab:pl}). Thus the magnetic field invalidates the Debye $T^3$
law; the notion of the Debye temperature cannot be applied to a
magnetized Coulomb crystal. At $b\gtrsim 1$ the saturation
temperature for branch 1 phonons scales as $1/b$. On the contrary,
the same growing magnetic field reduces $c_2(b,t)$ and increases the
power-law index (from 3 to 4) in quantum regime (Table \ref{tab:pl}).
In this case the magnetic field does not affect the saturation
temperature, and the heat capacity $c_2(b,t)$ at $b \gtrsim 1$
becomes almost independent of $b$. Accordingly, all curves for $b
\geq 1$ in the lower panel of Fig.\ \ref{f:capacity} merge, so that
when $b$ varies $c_2(b,t)$ actually changes only between the $b=0$
and $b=1$ curves. As for $c_3(t,b)$, the magnetic field increases the
saturation temperature. At $b \lesssim 1$ the increase is small (and
all the curves at $b<1$ almost merge), but at $b\gtrsim 1$ the
saturation temperature increases proportional to $b$. Below
saturation, $c_3(b,t)$ is exponentially small.

The upper panel of Fig.\ \ref{f:capacity} plots the total heat
capacity $c(b,t)$ (in natural scale) versus ${\rm log}_{10}t$ for the
same $b$. As long as $b\lesssim 1$, the magnetic field affects only
the low-temperature part of the $c(b,t)$-curve (which is invisible in
the upper panel of Fig.\ \ref{f:capacity} but which would be quite
visible in logarithmic scale, as in the lower panel --- see the left
panel of Fig.\ 2 in \citealt{B09}). A stronger field $b \gtrsim1$
dramatically changes $c(b,t)$ because of large separation of
saturation temperatures in branches 1, 2, and 3. With increasing $t$,
branch 1 saturates first at $t \sim 1/b$ and we have $c(b,t)\approx
1$; then branch 2 saturates at $t \sim 1$ after which $c(b,t)\approx
2$, and finally the last branch 3 saturates at $t \sim b$ giving
$c(b,t)\to 3$.

Figure \ref{f:energy} compares the temperature dependence of the heat
capacity $c(b,t)$ (thin lines) and internal energy $e(b,t)$ (thick
lines) for $b=$1, 10, 100, and 1000 (lines of different types).
Figure \ref{f:entropy} plots the free energy function $f(b,t)$ for
$b=$0, 10, 100, and 1000. As long as $b \lesssim 1$, the magnetic
field affects only low-temperature ($t \lesssim b$) parts of the
curves which are again invisible in natural scale of Figs.\
\ref{f:energy} and \ref{f:entropy} (cf.\ Figs.\ 2 and 4 in
\citealt{B09}). Accordingly, the functions $c(b,t)$, $e(b,t)$, and
$f(b,t)$  in our Figs.\ \ref{f:energy} and \ref{f:entropy}, if
plotted at several values of $b \lesssim 1$, would look almost the
same as those at $b=0$. However, at higher $b$ the magnetic field
strongly affects these thermodynamic functions. The temperature
dependence of $e(b,t)$ is seen to be smoother but similar to that of
$c(b,t)$: it reflects the saturation of different phonon branches at
different temperatures. Note that at $t\sim 1$ and any $b\gtrsim 1$
all functions $e(b,t)$ practically merge, and so do all functions
$c(b,t)$. This is because the behavior of these functions at $t \sim
1$ is determined by the branch 2 frequencies which cease to depend on
$b$ at $b \gtrsim 1$. The temperature dependence of $f(b,t)$ is
different because the free energy does not saturate but grows
logarithmically at classic temperatures.

After all the three phonon branches saturate at sufficiently high
temperature, total thermodynamic functions become independent of $b$
and coincide with those at $b=0$. In particular, $f(b,t)$ must be
given by the asymptote (\ref{e:asy:f0}). We have verified that our
numerical results are consistent with this expectation (cf.\ Table
\ref{tab:sat}).

\subsection{Zero-point contribution}
The Helmholz free energy and the internal energy
have the same extra (positive) contribution $(3 N/2) \varepsilon_0$ due to zero-point
ion vibrations. The mode-average oscillator energy $\varepsilon_0$ depends on $b$.
We can split $\varepsilon_0$ into three terms corresponding
to different phonon branches,
$\varepsilon_0 = \varepsilon_1+\varepsilon_2+\varepsilon_3$, where
$3 N \varepsilon_i = \sum_{\bm{k}} \hbar \Omega_{\bm{k}i}$, and summation is over the
first Brillouin zone.
We have calculated $\varepsilon_i$ and propose the following analytical expressions
to describe the numerical results:
\begin{eqnarray}
       \varepsilon_1 &=& \frac{0.09005 \,\, \hbar \omega_{\rm p}}{\sqrt{1+0.3981 b+1.795 b^2}}~,
\nonumber \\
      \frac{\varepsilon_2}{\hbar \omega_{\rm p}} &=& \frac{0.1379}{1+4.542 b^2}+\frac{0.1804 b^2}{b^2+1/4.542}~,
\nonumber \\
    \varepsilon_3 &=& \hbar \omega_{\rm p} \sqrt{0.2834^2+b^2/9}~.
\label{moments}
\end{eqnarray}
The rms errors of these fits are below 1\%. Maximum errors are below
2\%. At $b=0$ we have $\varepsilon_0/\hbar \omega_{\rm p} \approx 0.5114$, which is the
well-known moment $u_1$ of the bcc lattice \citep[][]{C61}.
At $b \gg 1$ the dominant
contribution is due to the third branch: $\varepsilon_3 \approx \hbar \omega_{\rm B}/3$.

\section{Discussion}
\label{s:discuss}
%

%******************************************************************
%
\begin{figure}
\vspace{-0.5cm}
\begin{center}
\leavevmode
\includegraphics[width=74mm]{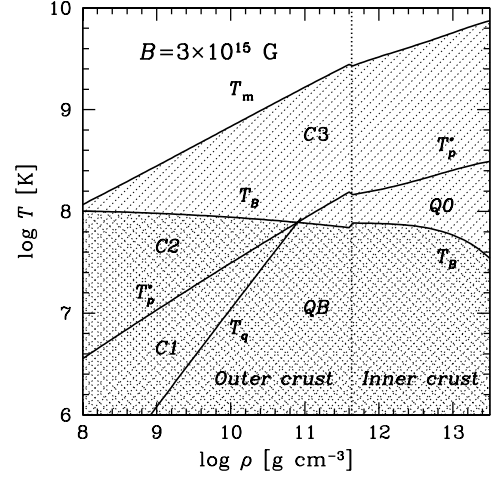}
\end{center}
\caption[ ]{$T-\rho$ regions at $B=3 \times 10^{15}$ G in the neutron
star crust composed of ground-state matter where heat capacity of
Coulomb crystal has different behavior (Table \ref{tab:cregimes}).
$T_\mathrm{m}$ is the melting temperature; vertical dotted line
separates the outer and inner crust. In the singly-shaded region the
$B$-field effects are unimportant, while in the doubly-shaded region
they affect the heat capacity. See text for details.}
\label{f:diagb3e15}
\end{figure}
%
%******************************************************************

\begin{table}
\caption[]{Asymptotic behavior of total crystal heat capacity per one
ion in different $T-\rho$ regions shown in Fig.\ \ref{f:diagb3e15}}
\label{tab:cregimes}
\begin{center}
\begin{tabular}{c c c c  }
\hline \hline
Regime & Dominant branch & $B$-field effect & Asymptote of $c$ \\
\hline
{\it C3} & 1,2,3  & absent & 3 \\
{\it C2} & 1,2 &  present & 2  \\
{\it C1} & 1   &  present & 1  \\
\hline  {\it Q0}  & 1,2 & absent & $2513\, t^3$ \\
{\it QB} & 1 & strong & $31\, (bt)^{3/2}$ \\
\hline \hline
\end{tabular}
\end{center}
\end{table}

%******************************************************************
%                                                       entropy
\begin{figure*}
%\vspace{-0.5cm}
\begin{center}
\leavevmode
\includegraphics[width=170mm]{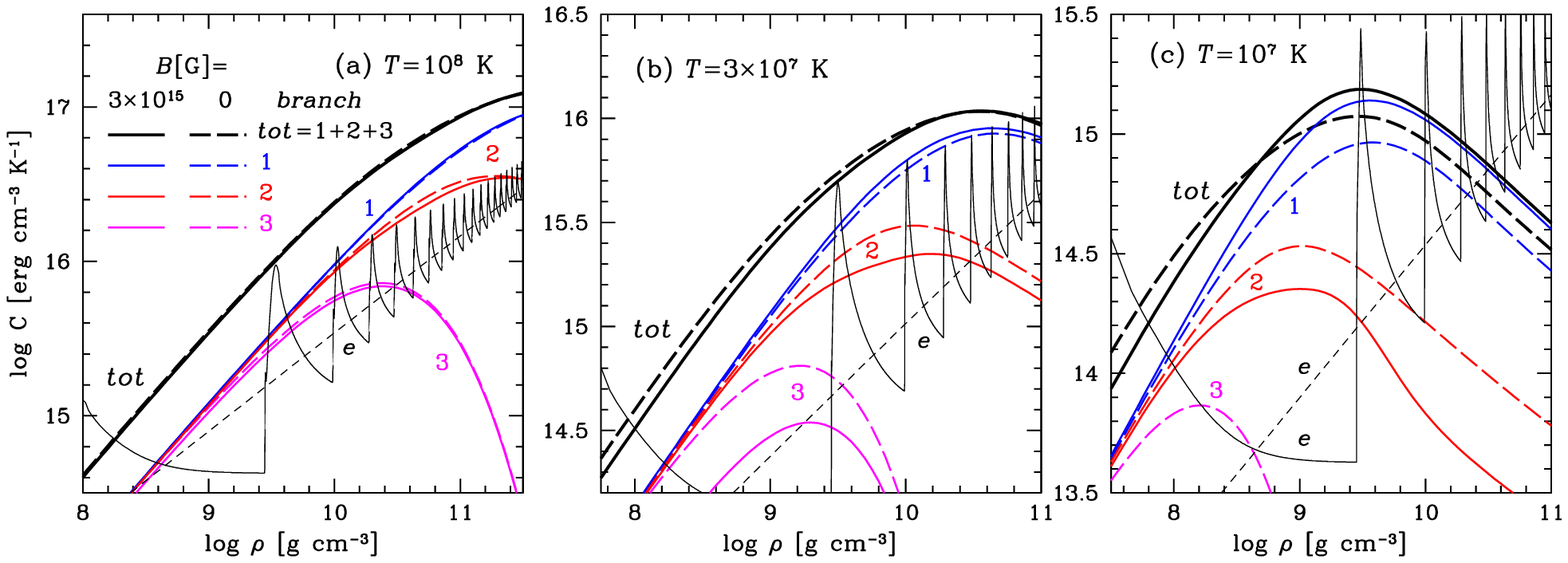}
\end{center}
\caption[ ]{(Color online) Specific heat capacity (at constant volume
per cubic centimeter) versus density in the outer crust of a neutron
star for $B=3\times 10^{15}$ G (solid lines) and $B=0$ (dashed
lines). Three panels (a), (b), and (c) are for $T=10^8$, $3 \times
10^7$, and $10^7$ K, respectively. Medium-width lines 1, 2, and 3
show partial crystal heat capacities due to respective phonon
branches. Thick lines {\it tot} are total crystal heat capacities.
Thin lines {\it e} are electron heat capacities. See text for
details.} \label{f:capb3e15}
\end{figure*}
%
%******************************************************************

In this section we apply the analytical formulas of Sect.\
\ref{s:fit} to study the heat capacity of the magnetized crystalline
neutron star crust. The temperature-density diagram of the crust with
$B=3\times10^{15}$~G is plotted in Fig.\ \ref{f:diagb3e15}. We assume
the ground-state composition of the crust and use the simplified
smooth-composition model \citep{HPY07}. The same diagram for the
accreted crust composition (e.g., \citealt{HPY07}) should be
qualitatively similar. For simplicity, we neglect magnetic field
effects on the nuclear composition of the crust.

The dotted vertical line in Fig.\ \ref{f:diagb3e15} shows the neutron
drip point ($\rho_\mathrm{ND}\approx 4.3 \times 10^{11}$
g~cm$^{-3}$). It separates the outer and inner crust. The outer crust
consists of electrons and ions; the latter are fully ionized by the
electron pressure for rather high densities displayed in the figure.
The inner crust consists of electrons, ions and free neutrons. The
inner crust extends to the density $\approx 1.5 \times 10^{14}$ g
cm$^{-3}$. We do not display the highest-density inner crust because
its composition is not very certain (may contain funny pasta phases
of nuclear clusters, as reviewed, for instance, by \citealt{HPY07}).

The upper line in Fig.\
\ref{f:diagb3e15} is the melting temperature $T_{\rm m}$ of the
field-free crystal.  Our analysis is thus limited to the shaded
region below $T_{\rm m}$.
Various other lines in Fig.\ \ref{f:diagb3e15}
split the $T-\rho$ plane into
several domains ($C1$, $C2$, $C3$, $Q0$, and $QB$), where the ion heat
capacity shows qualitatively different behavior at
$B=3\times 10^{15}$~G. These regimes are listed in
Table \ref{tab:cregimes}. Notice that the boundaries between the domains
are approximate and the change of heat capacity when moving
from one domain to another is smooth.

The line $T_\mathrm{B}$ is the ion cyclotron temperature (defined as
$T_\mathrm{B}=\hbar \omega_\mathrm{B}$ and expressed in Kelvins). At
$T>T_\mathrm{B}$ the magnetic field has almost no effect on the heat
capacity of ions (domains $C3$ and $Q0$). In the doubly-shaded region
$T<T_\mathrm{B}$ (domains $C2$, $C1$, and $QB$) the magnetic field
affects the ion heat capacity. For $B=3\times 10^{15}$~G the ion heat
capacity of the crust is thus affected as long as $T \lesssim
10^8$~K. Straightforward scaling implies that the field $B \sim 3
\times 10^{14}$ G becomes important at $T \lesssim 10^7$~K, while
$B\sim 3 \times 10^{16}$ G modifies the ion heat capacity at $T
\lesssim 10^9$~K.

The temperature $T_\mathrm{p}^*$ in Fig.\ \ref{f:diagb3e15} is
defined as $T_\mathrm{p}^*= 0.1\,T_\mathrm{p}$. It is a better
measure of quantum effects in a $B=0$ Coulomb crystal than
$T_\mathrm{p}$ itself. We also plot the temperature
$T_\mathrm{q}=T_\mathrm{p}^{*2}/T_\mathrm{B}$ (at those mass
densities $\rho$, where $T_\mathrm{q}<T_\mathrm{p}^*$). As $B$
increases, this line shifts to the right ($T_\mathrm{q}$ becomes
lower at given $\rho$).

Domain $C3$ in Fig.\ \ref{f:diagb3e15} corresponds to classic
(high-temperature) crystal where all the three phonon branches are
saturated and the magnetic field does not affect the heat capacity
($c\approx3$, Table \ref{tab:cregimes}). In domain $C2$ the magnetic
field reduces the contribution of phonon branch 3, but branches 1 and
2 are still saturated ($c \approx 2$). In domain $C1$ the magnetic
field reduces the contribution of phonon branch 2 although branch 1
remains saturated ($c \approx 1$). When $T$ falls below
$T_\mathrm{q}$ one enters domain $QB$, which corresponds to quantum
power law regime IV.1 of Fig.\ \ref{f:modes}. All phonon branches are
non-saturated, and the leading heat capacity $\propto T^{3/2}$ is
produced by branch 1. The magnetic field effect is the strongest in
this regime. The variation of the heat capacity in these movings
($C3\to C2\to C1\to QB$) can be easily understood from Fig.\
\ref{f:capacity}. At $T_\mathrm{p}^* > T > T_\mathrm{B}$ (domain
$Q0$) one has a quantum crystal unaffected by magnetic field. The ion
heat capacity there is $\propto T^3$ and is due to phonon branches 1
and 2 (regimes I.1 and I.2 in Fig.\ \ref{f:modes}).

Let us mention that for any value of $B$ from $\sim 10^{14}$~G to
$\sim 3 \times 10^{16}$~G, one can estimate the ion heat capacity in
the crust by rescaling the lines $T_\mathrm{B}$ and $T_\mathrm{q}$ in
Fig.\ \ref{f:diagb3e15}. At $B \lesssim 10^{14}$~G the temperature
$T_\mathrm{B}$ becomes lower than $T_\mathrm{p}^*$ and at $B \gtrsim
3 \times 10^{16}$~G it becomes higher than $T_\mathrm{p}^*$ in the
entire density range displayed, and some domains move out of the
picture.

The above discussion is further illustrated in Fig.\
\ref{f:capb3e15}. It compares ion and electron heat capacities (at
constant volume per cubic centimeter) as functions of density in the
outer neutron star crust. Three panels (a), (b), and (c) are for
$T=10^8$, $3 \times 10^7$ and $10^7$ K, respectively. Solid and
dashed lines represent heat capacities at $B=3 \times 10^{15}$~G and
at $B=0$. Thick lines (labeled as {\it tot}) plot the total ion heat capacity
(sum over three phonon branches). Medium-thick lines (labeled as 1,
2, and 3) display the partial contributions of phonon branches 1, 2,
and 3, respectively. Finally, thin lines
(denoted as $e$)
present the electron heat capacities.  The electron heat capacity in
a magnetic field (thin solid line) oscillates with growing density
because degenerate electrons populate new Landau levels. With
decrease of temperature, the oscillations become more pronounced.
Comparing the ion and electron heat capacities in Fig.\
\ref{f:capb3e15}, we conclude that the ion heat capacity, for the
most part, dominates in the outer crust for the given temperature
range.

The behavior of the ion heat capacity in Fig.\ \ref{f:capb3e15} is
easily understood from Fig.\ \ref{f:diagb3e15}. The temperature
$T=10^8$~K in Fig.\ \ref{f:capb3e15}(a) is close to $T_\mathrm{B}$.
Accordingly, the field $B=3\times10^{15}$~G affects the ion heat
capacities provided by all phonon branches 1, 2 and 3 only weakly
(so that ion solid and dashed lines almost merge).
A visible reduction of these heat capacities at higher $\rho$ is
explained by the transition from classic to quantum ion crystal. With
the drop of $T$ [in Figs.\ \ref{f:capb3e15}(b) and (c)] the effect of
the magnetic field on the ion heat capacity becomes stronger. Quantum
suppression of the ion heat capacity also grows stronger. These
results are in accord with Fig.\ \ref{f:diagb3e15}. For instance, the
phonon branch 3 contribution at $B=3\times 10^{15}$~G and $T=10^7$ in
Fig.\ \ref{f:capb3e15}(c) is so small that it is not shown in the
figure.

Summarizing, one can say that magnetic field $B \sim 3\times10^{15}$~G
may have an important effect on the ion heat capacity in the outer crust.
The ion heat capacity in the inner crust will be also modified by such
magnetic field but it will happen at those temperatures
where the electron heat capacity dominates
[cf.\ Fig.\ \ref{f:capb3e15}(c)]. For lower $B$, the magnetic field
affects the ion heat capacity in smaller ranges of $T$ and $\rho$ in
the outer crust.

%******************************************************************
%
\begin{figure}
\vspace{-0.5cm}
\begin{center}
\leavevmode
\includegraphics[width=76mm]{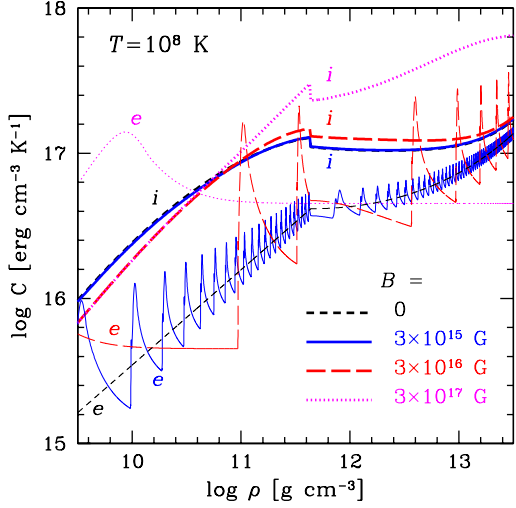}
\end{center}
\caption[ ]{(Color online) The total ion (thick lines $i$) and
electron (thin lines $e$) heat capacities for $T=10^8$~K versus
density in the outer and inner neutron star crust at $B=0$ (short
dashes), $3\times10^{15}$~G (solid lines), $3\times10^{16}$~G (long
dashes) and $3\times10^{17}$~G (dotted lines). } \label{f:cbt8}
\end{figure}
%
%******************************************************************

Larger fields $B > 3 \times 10^{15}$~G can influence the ion heat
capacity in the outer and inner crust at higher $T \gtrsim 10^8$~K.
To illustrate this statement, in Fig.\ \ref{f:cbt8} we plot the total
ion (thick lines) and electron (thin lines) heat capacities for
$T=10^8$~K versus density in the outer and inner crust at $B=0$
(short-dashed lines) and three values of $B$, $3 \times 10^{15}$~G
(solid lines), $3 \times 10^{16}$~G (long-dashed lines), and $3
\times 10^{17}$~G (dotted lines). The lines for $B=0$ and $3 \times
10^{15}$~G are essentially the same as in Fig.\ \ref{f:capb3e15}(a),
but now they are extended to the inner crust. Some jumps of the ion
heat capacity occur at the neutron drip point. We see that the field
$B \gtrsim 3\times 10^{16}$~G, indeed, noticeably enhances the ion
heat capacity in the inner crust, and this contribution will dominate
over the electron one. Note a delay of quantum oscillations of the
electron heat capacity after the neutron drip point (at $\rho \gtrsim
\rho_\mathrm{NB}$, $B=3 \times 10^{15}$ and $3 \times 10^{16}$ G). It
is due to the efficient neutronization of matter which is accompanied
by a slower growth of the electron number density (and of the
electron chemical potential that regulates the quantum oscillations)
with increasing $\rho$. The quantum oscillations just after the
neutron drip are very sensitive to the density dependence of the
electron number density.

%******************************************************************
%
\begin{figure}
\vspace{-0.5cm}
\begin{center}
\leavevmode
\includegraphics[width=76mm]{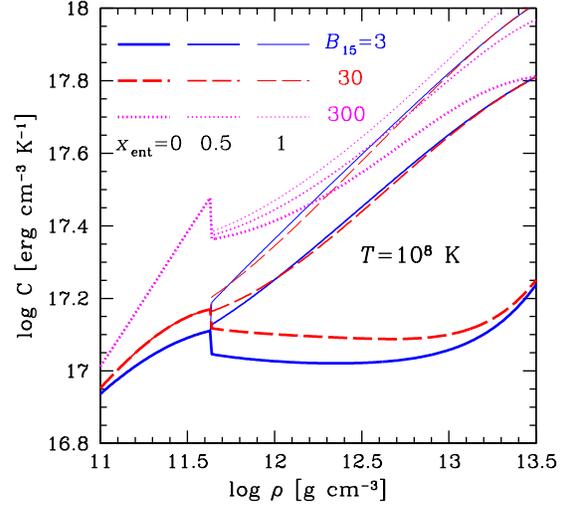}
\end{center}
\caption[ ]{(Color online) The total ion heat capacity for $T=10^8$~K
versus density in the neutron star crust at $B=3\times10^{15}$~G
(solid lines), $3\times10^{16}$~G (long dashes) and
$3\times10^{17}$~G (dotted lines). Thick lines correspond to the
standard ground state matter; thinner lines are for the matter where
one halve of free neutrons are entrained to the nuclei; even thinner
lines -- all neutrons are entrained (see text for details).}
\label{f:cbt8cham}
\end{figure}
%
%******************************************************************

An order of magnitude higher magnetic field $B = 3\times 10^{17}$~G
would exert the most drastic influence on the ion heat capacity of
the entire inner crust of a neutron star. Under conditions shown in
Fig.\ \ref{f:cbt8} the ion heat capacity at $B = 3\times 10^{17}$~G
exceeds the field-free value by up to a factor of 4. The reason for
this behavior can be traced back to Fig.\ \ref{f:capacity}, where the
branch 1 heat capacity of a quantum crystal is seen to be greatly
amplified by the magnetic field. The strongest effect may be expected
at temperatures $T \lesssim 0.1 T_{\rm p}^*$ (Fig.\
\ref{f:diagb3e15}), that is at $T \lesssim 0.01 T_{\rm p} \approx 6
\times 10^7 Z_{40} \sqrt{\rho_{14} X/4} /A_{1e3}$ K, combined with $b
\gtrsim 1$ or $B \gtrsim 5 \times 10^{17} \sqrt{4 \rho_{14}/X}$ G. In
this case $\rho_{14}$ is the mass density in units of $10^{14}$ g
cm$^{-3}$, $Z_{40} = Z/40$, $A_{1e3} = A/1000$, where $A$ is the
number of nucleons per one nucleus (including free neutrons in the
inner neutron star crust), and $X=A/A_{\rm N}$, where $A_{\rm N}$ is
the number of nucleons bound in a nucleus ($A_{\rm N}$ determines ion
plasma and cyclotron frequencies).

The field $B=3 \times 10^{17}$~G is so large that plasma electrons
occupy the ground Landau level throughout the whole crust.
The heat capacity (per
cm$^3$) of relativistic degenerate electrons is then independent of
the electron number density (and hence independent of $\rho$).
The quantum oscillations of the electron heat capacity are absent.
The bump of the electron heat capacity at $\rho \sim 10^{10}$ is
caused by the onset of electron degeneracy.

Let us remark that the heat
capacity in the inner crust contains also the contribution of free
neutrons (omitted in this work). For non-superfluid neutrons, this contribution would
dominate. However, it is strongly suppressed and becomes small
if neutrons are superfluid \citep[e.g.,][]{GYP01}.

Finally, Fig.\ \ref{f:cbt8cham} illustrates the effects of neutron
entrainment in the inner crust on the heat capacity of ions. It shows
the total ion heat capacity versus density at $T=10^8$ K for the same
values of $B=3 \times 10^{15}$, $3 \times 10^{16}$ and $3 \times
10^{17}$ G as in Fig.\ \ref{f:cbt8} (solid, dashed and dotted lines,
respectively). The thick lines are for the same standard model of the
inner crust as in Fig.\ \ref{f:cbt8}. Thinner lines take into account
recent theoretical predictions (e.g., \citealt{cham12},
\citealt{cham13}) that some fraction of free neutrons can be actually
entrained by the atomic nuclei. We have incorporated this effect in a
schematic manner by increasing masses of atomic nuclei in the inner
crust. This reduces the ion plasma and cyclotron frequencies  and the
suppression of the heat capacity in the quantum regime. The medium
size lines are calculated assuming that one half of the free neutrons
are entrained in this way ($x_{ent}=0.5$), while the thinnest lines
assume that all free neutrons are entrained ($x_{ent}=1$). We see
that the entrainment effect can strongly enhance the ion heat
capacity in the inner crust.

The present results are important for modeling the structure,
evolution and observational manifestations of neutron stars with
strong crustal magnetic fields such as magnetars
(SGRs and AXPs) and high magnetic field pulsars (Sect.\ \ref{introduct}).
These objects are observationally linked to a wide range of exciting
astrophysical phenomena.
The internal crustal magnetic fields of these objects are thought to have both, poloidal
and toroidal, components ($B_\mathrm{p}$ and $B_\mathrm{tor}$), and
both components can be substantial,
especially in the beginning of neutron star lives. For instance,
\citet{pp11} simulated the magneto-thermal evolution of the magnetar
(AXP) 1E 2259+586 assuming the initial magnetic field values $B_p=2.5
\times 10^{14}$~G and $B_\mathrm{tor}=2.5 \times 10^{16}$~G.
Our results should help modeling magnetars, high-$B$
pulsars, and their possible mutual transformations (Sect.\ \ref{introduct}).
Primarily, we mean modeling thermo-magnetic
evolution of these objects, dynamics of giant and ordinary bursts in
magnetars and afterburst relaxation.

\section{Conclusions}
We have performed accurate calculations of the phonon thermodynamic
functions of the magnetized bcc Coulomb crystal and approximated our
numerical results by analytic expressions. Thermodynamic properties
are fully determined by the free energy function $f(b,t)$ of
dimensionless temperature $t$ and magnetic field $b$ [see Eqs.\
(\ref{e:thermodyn}) and (\ref{e:thermodyn_i})] which we split into
three functions $f(b,t)=f_1(b,t)+f_2(b,t)+f_3(b,t)$ for the three
phonon branches. We have derived three fit expressions (\ref{e:f1b}),
(\ref{e:f2}), and (\ref{e:f3}) for the partial functions $f_i(b,t)$.
For a non-magnetized crystal the partial functions $f_i(0,t)$ are
given by simpler fit expressions (\ref{e:f120}) and (\ref{e:f30}).
Our analytic fits are sufficiently simple for differentiating and
obtaining other thermodynamic functions without any significant loss
of accuracy (Table \ref{tab:errors}). In this way we derive
selfconsistent analytic description of harmonic-lattice
thermodynamics. A similar (simpler but less accurate) description has
been obtained earlier by \citet{pc13}. Although our calculations have
been performed for restricted values of $t$ and $b$ ($10^{-4} \leq t
\leq 10^{4}$; $b=0$ and $10^{-3} \leq b \leq 10^3$) we expect that
the analytic fits remain accurate for wider ranges of $t$ and $b$. We
have also analyzed various behaviors of the thermodynamic functions
(Tables \ref{tab:modes}--\ref{tab:sat}) which may be used for
estimating these functions under specific conditions. The results are
obtained for one specific orientation of the magnetic field in the
crystal but the dependence of thermodynamic functions on the
orientation is weak \citep{B00,B09} and is thus unimportant for many
applications.

In addition, we have analyzed (Sect.\ \ref{s:discuss}) the behavior
of the heat capacity of crystallized ions in the crust of a strongly
magnetized neutron star. We have compared the ion heat capacity with
the electron one. We have shown that the field $B \lesssim 3 \times
10^{15}$~G affects the ion heat capacity in the outer crust at $T
\lesssim 10^8$~K; in this case the ion heat capacity mainly dominates
over the electron one. Stronger $B$-fields can affect the ion heat
capacity both in the inner and outer crusts and at higher $T$.
Moreover, we have demonstrated that the ion heat capacity in the
inner crust is sensitive to the effect of free neutron entrainment by
atomic nuclei (\citealt{cham12}, \citealt{cham13}). We believe that
our results will be helpful for modeling observational manifestations
of magnetars and high magnetic field pulsars.

\section*{Acknowledgments}
We are grateful to A.Y. Potekhin for constructive criticism of the
initial version of this paper.
The work was supported by Ministry of Education and Science of 
the Russian Federation (Agreement No. 8409), 
by RFBR (grant 11-02-00253-a), and by Rosnauka (grant NSh 4035.2012.2).

\label{lastpage}


\begin{thebibliography}{}

\bibitem[\protect\citeauthoryear{Albers \& Gubernatis}{1981}]{AG81}
    Albers R.C., Gubernatis J.E., 1981, Los Alamos Scientific
    Laboratory Report No.\ LA-8674-MS

\bibitem[\protect\citeauthoryear{Baiko}{2000}]{B00}
    Baiko D.A., 2000, PhD thesis,
    A.F. Ioffe Physical-Technical Institute

\bibitem[\protect\citeauthoryear{Baiko, Potekhin \& Yakovlev}{Baiko et al.}{2001}]{BPY01}
      Baiko D.A., Potekhin A.Y., Yakovlev D.G., 2001,
      Phys. Rev. E, 64, 057402

\bibitem[\protect\citeauthoryear{Baiko}{2009}]{B09}
      Baiko D.A., 2009, Phys. Rev. E, 80, 046405

\bibitem[\protect\citeauthoryear{Carr}{1961}]{C61}
      Carr W.J., 1961, Phys. Rev., 122, 1437

\bibitem[\protect\citeauthoryear{Chamel}{Chamel}{2013}]{cham13}
     Chamel N., 2013,  Phys. Rev. Lett. 110, 011101

\bibitem[\protect\citeauthoryear{Chamel}{Chamel, Page \& Reddy}{2013}]{cham12}
     Chamel N., Page D., Reddy S., 2013, Phys. Rev. C 87, 035803

\bibitem[\protect\citeauthoryear{Dubin \& O'Neil}{Dubin \& O'Neil}{1999}]{dubin99}
     Dubin D. H. E., O'Neil T. M., 1999, Rev. Mod. Phys. 71, 87

\bibitem[\protect\citeauthoryear{Gnedin, Yakovlev \& Potekhin}{Gnedin et al.}{2001}]{GYP01}
    Gnedin O.Y., Yakovlev D.G., Potekhin A.Y., 2001,
    MNRAS, 324, 725

\bibitem[\protect\citeauthoryear{Haensel, Potekhin \& Yakovlev}{2007}]{HPY07}
     Haensel P., Potekhin A.Y., Yakovlev D.G., 2007,
     Neutron Stars 1: Equation of State and Structure. Springer, New York

\bibitem[\protect\citeauthoryear{Itano et al.}{Itano et al.}{1998}]{itano98}
    Itano W. M., Bollinger J. J.,  Tan J. N., Jelenkovic B.,  Huang X.-P.,
    and Wineland D. J., 1998, Science 279, 686

\bibitem[\protect\citeauthoryear{Kittel}{1995}]{K95}
      Kittel C., 1995, Introduction to Solid State Physics, 7th edition.
      Wiley

\bibitem[\protect\citeauthoryear{Landau \& Lifshitz}{1980}]{LL80}
      Landau L.D., Lifshitz E.M., 1980, Statistical Physics. Part I.
      Pergamon Press, Oxford

\bibitem[\protect\citeauthoryear{Livingstone et al.}{Livingstone et al.}{2011}]{livingstone11}
     Livingstone M. A., Ng C.-Y., Kaspi V. M., Gavriil F. P., Gotthelf E.
     V., 2011,  Astrophys. J. 730, 66

\bibitem[\protect\citeauthoryear{Mereghetti}{Mereghetti}{2008}]{sandro08}
     Mereghetti S., 2008, Annual Rev. Astron. Astrophys. 15, 225

\bibitem[\protect\citeauthoryear{Nagai \& Fukuyama}{Nagai \& Fukuyama}{1982}]{NF82}
     Nagai T., Fukuyama H., 1982,
     J.\ Phys.\ Soc.\ Jap., 51, 3431

\bibitem[\protect\citeauthoryear{Nagai \& Fukuyama}{Nagai \& Fukuyama}{1983}]{NF83}
      Nagai T., Fukuyama H., 1983,
      J.\ Phys.\ Soc.\ Jap., 52, 44

\bibitem[\protect\citeauthoryear{Pollock \& Hansen}{Pollock \& Hansen}{1973}]{PH73}
     Pollock E.L., Hansen J.P., 1973,
     Phys. Rev. A, 8, 3110

\bibitem[\protect\citeauthoryear{Pons, Miralles \& Geppert}{Pons et al.}{2009}]{pmg09}
     Pons J. A., Miralles J. A., Geppert U., 2009,  Astron. Astrophys.
     496, 207

\bibitem[\protect\citeauthoryear{Pons \& Perna}{Pons \& Perna}{2011}]{pp11}
     Pons J. A., Perna R., 2011, Astrophys. J. 741, 123

\bibitem[\protect\citeauthoryear{Potekhin \& Chabrier}{Potekhin \& Chabrier}{2013}]{pc13}
     Potekhin A. J., Chabrier G., 2013,  Astron. Astrophys.
     550, A43

%\bibitem[\protect\citeauthoryear{Shapiro \& Teukolsky}{1983}]{ST83}
%       Shapiro S.L., Teukolsky S.A., 1983, Black Holes, White Dwarfs
%       and Neutron Stars: The Physics of Compact Objects.
%       Wiley-Interscience

\bibitem[\protect\citeauthoryear{Usov, Grebenschikov \& Ulinich}{Usov et al.}{1980}]{UGU80}
      Usov N.A., Grebenschikov Yu.B., Ulinich F.R., 1980,
      J. Exp. Theor. Phys. 78, 296

\bibitem[\protect\citeauthoryear{Woods \& Thompson}{2006}]{WT06}
    Woods P.M., Thompson C., 2006, in: Compact stellar X-ray sources,
    eds.\ W. Lewin and M.\ van der Klis, Cambridge University Press, Cambridge

\bibitem[\protect\citeauthoryear{Zhu et al.}{Zhu et al.}{2011}]{zhuetal11}
     Zhu W. W., Kaspi V. M., McLaughlin M. A., Pavlov G. G., Ng C.-Y.,
     Manchester R. N., Gaensler B. M., Woods P. M., 2011, Astrophys. J.
     734, 44

\end{thebibliography}
\end{document}